\documentclass[fleqn]{article}
\usepackage{multicol}
\usepackage{graphicx} 
\usepackage{geometry}
\usepackage[utf8]{inputenc}
\usepackage[T1]{fontenc}
\usepackage{setspace}
\usepackage{amsmath}
\usepackage{bbm}
\usepackage{float}
\usepackage{microtype}
\usepackage{amssymb}
\usepackage{amsthm}
\usepackage{physics}

\usepackage{tikz}
\usepackage{hyperref}
\usepackage{pgf-pie}
\usepackage{graphicx}
\usepackage{pdfpages}
\usepackage{color}
\usepackage{mathtools}
\usepackage{enumerate}
\usepackage[inkscapelatex=false]{svg}
\usepackage[OT1]{fontenc}
\usepackage[
    backend=biber,
    style=nature]{biblatex} 
\usepackage{authblk}
\usepackage{enumitem}

\newcommand{\Okl}{\Omega_{k,l}}

\newcommand{\E}{\mathcal{E}}

\newcommand{\mP}{\mathcal{P}}
\newcommand{\mQ}{\mathcal{Q}}

\newcommand{\Zd}{\mathcal{Z}_d}
\newcommand{\w}{\omega}
\newcommand{\denop}{\mathcal{D}}
\newcommand{\hs}{\mathcal{H}}

\newcommand{\otAB}{\overset{AB}{\otimes}}

\newcommand{\prob}{\mathbbm{p}}
\newcommand{\Prob}{\mathbbm{P}}

\title{Low-Fidelity Entanglement Distillation with FIMAX}
\author[1]{Christopher Popp}
\author[2]{Tobias C. Sutter}
\author[3]{Beatrix C. Hiesmayr}
\affil[1,2,3]{University of Vienna, Faculty of Physics, Währingerstrasse 17, 1090 Vienna.\vspace{3.5mm}}
\affil[1]{christopher.popp@univie.ac.at}
\affil[2]{tobias.christoph.sutter@univie.ac.at}
\affil[3]{beatrix.hiesmayr@univie.ac.at}
\date{}
\begin{document}
\maketitle
\onehalfspacing
\begin{abstract}
Uncontrolled interactions with the environment introduce errors that remain a significant challenge to the reliability of quantum technologies using entanglement. An essential method to overcome or mitigate these errors is entanglement distillation, the transformation of multiple copies of weakly entangled states into a smaller number of approximately maximally entangled states. We present a comparative analysis of the stabilizer-based two-copy entanglement distillation protocol, FIMAX, against other recurrent two-copy protocols, including ADGJ, DEJMPS, P1-P2, and the generalized BBPSSW protocol. We focus on low-fidelity bipartite quantum states in dimensions $d=2$ and $d=3$, which are particularly challenging to distill. Our findings demonstrate that FIMAX exhibits superior performance for these states. While other protocols struggle with highly noisy states, FIMAX successfully distills entanglement even when the initial state quality is severely compromised. These results highlight the protocol's capability to address the effects of environmental noise, advancing the robustness and scalability of quantum technologies leveraging entanglement distillation.
\end{abstract}
\begin{multicols}{2}
\section{Introduction}
Overcoming the pervasive challenge of environment-induced errors is crucial for quantum technologies to revolutionize applications like secure communication \cite{pirandola_advances_2020} and computation \cite{wang_qudits_2020}. \\
Quantum information processing tasks often require quantum-specific resources, such as entanglement and coherence, to function reliably \cite{chitambar_quantum_2019, bauml_resource_2019}. Affecting the entanglement structure \cite{popp_special_2024}, noise in prepared quantum states, operations, or measurements can degrade these resources and, consequently, the performance of reliant applications. Even in the fundamental case of a bipartite qudit system, a $d^2$-dimensional quantum system shared between two parties, the correction or mitigation of errors is essential for realizing tasks on Noisy Intermediate-Scale Quantum (NISQ) devices \cite{preskill_quantum_2018, acharya_quantum_2024,bravyi_high-threshold_2024}. To tackle this challenge, two promising approaches have emerged: Quantum error correction \cite{steane_error_1996} and entanglement distillation \cite{bennett_mixed-state_1996}. Error correction involves encoding logical information into states resilient to specific forms and levels of noise, enabling the detection and subsequent correction of errors. Entanglement distillation aims to extract pure, maximally entangled states from noisy shared states through local operations and classical operations, thereby facilitating reliable implementation of various quantum information processing tasks. Notably, these approaches are interconnected \cite{dur_entanglement_2007}: Any quantum error-correcting code can be utilized in entanglement distillation schemes, and conversely, maximally entangled states produced by any distillation protocol enable reliable quantum communication, e.g., via teleportation \cite{bennett_teleporting_1993}.
Recently, a highly effective entanglement distillation protocol named FIMAX \cite{popp_novel_2024} has been proposed, leveraging this relationship. Based on stabilizer codes, a class of quantum error correction codes \cite{gottesman_stabilizer_1997, matsumoto_conversion_2003}, FIMAX exploits algebraic relations between these codes and maximally entangled states to efficiently use noisy bipartite quantum states for distillation. This protocol demonstrates significant promise in enhancing the fidelity, i.e., the overlap of used states with the maximally entangled state, even if the initial states are substantially affected by noise. \\ 
In this work, we perform a comprehensive analysis of the distillation capabilities of FIMAX and compare it to other recurrent entanglement distillation protocols that operate on two copies of the input state. We focus on entangled bipartite qudit states subjected to very high levels of noise, yet still exhibiting entanglement. Unlike hashing or breeding protocols \cite{bennett_mixed-state_1996, vollbrecht_efficient_2003}, which typically require states with higher fidelities, recurrent distillation protocols like FIMAX can  distill very noisy states, although their effectiveness is influenced by the specific characteristics of the noise. Note that there exist entangled, yet generally undistillable states, so-called bound entangled states \cite{horodecki_mixed-state_1998} (for a recent review see Ref.~\cite{hiesmayr_bipartite_2024}).  The proportion of entangled low-fidelity states that can be successfully distilled depends on the protocol applied, and is the main quantity of interest in this work. \\
The paper is organized as follows: In Sec.~\ref{sec:preliminaries}, we introduce entanglement distillation using FIMAX, emphasizing the key concepts necessary to understand its principles and subsequent analyses. Sec.~\ref{sec:results} presents our main results, comparing FIMAX to other distillation protocols regarding the share of distillable states across various families of low-fidelity bipartite states. We also report several interesting observations about FIMAX's distillation properties. Sec.~\ref{sec:conclusion} concludes our numerical analysis and the implications of our findings for future quantum technologies. In Sec.~\ref{sec:methods}, we detail the technical aspects and methods concerning the analyzed error-affected states, the stabilizer theory underpinning FIMAX, and the protocol itself. Additionally, we provide a simple example to enhance comprehension of these techniques. 
\section{Preliminaries}
\label{sec:preliminaries}
In this section, we introduce the basic notation for the application of the entanglement distillation protocol \mbox{FIMAX}~\cite{popp_novel_2024} and leave the details to Sec.~\ref{sec:methods}. 
The goal of entanglement distillation protocols is to create maximally entangled states from potentially many copies of the generally error-affected mixed input state $\rho_{in}$ that is shared between two parties named Alice and Bob. For this task, only local operations and classical communication (LOCC) are allowed to be used for Alice and Bob.
We choose the specific maximally entangled $d^2$-dimensional bipartite state
\begin{flalign}
    \label{eq:omega00}
    |\Omega_{0,0}\rangle &:= \frac{1}{\sqrt{d}}\sum_{i=0}^{d-1} | i \rangle \otAB |i \rangle 
\end{flalign}
to be the target state. This state is one of the so-called Bell states, denoted by $|\Omega_{k,l}\rangle$ and defined in Sec.~\ref{sec:bellstates}. Note that multiple copies of this target state with $d>2$ can always be transformed to the often considered maximally entangled qubit state $|\Phi^+\rangle := \frac{1}{\sqrt{2}} \sum_{i=0}^1 |i\rangle \otAB |i\rangle$ ($~\equiv |\Omega_{0,0}\rangle$ for $d=2$) via LOCC. Hence, any protocol that produces the states \eqref{eq:omega00} for $d > 2$ can always be extended to distill $|\Phi^+\rangle$ in a lower-dimensional subspace. \\
We shortly describe the principles of the two-copy stabilizer-based recurrent entanglement distillation protocol, FIMAX. 
The protocol consists of the following main steps:
First, Alice and Bob transform two copies of the input state via LOCC into Bell-diagonal form, in which states can be interpreted as mixtures of Bell states that are affected by so-called Weyl errors.
Second, Alice and Bob perform local measurements using both copies of their states. Thereby, they choose a specific measurement basis, which is determined by the properties of the input state and related to a certain stabilizer, a subgroup of the operators representing the Weyl errors.
Finally, they perform local operations based on the stabilizer and their measurement outcomes. 
If successful, the protocol results in an output state $\rho_{out}$, having a higher fidelity with the target state, i.e., $\mathcal{F}_{out} = \langle \Omega_{0,0} | \rho_{out} | \Omega_{0,0} \rangle \geq \langle \Omega_{0,0} | \rho_{in} | \Omega_{0,0}  \rangle = \mathcal{F}_{in}$.
\mbox{FIMAX} has been implemented as part of the open source Julia package ``BellDiagonalQudits.jl'' \cite{popp_belldiagonalqudits_2023}.\\ 
We have summarized the details regarding the relevant theory and application of the protocol in Sec.~\ref{sec:methods}. The Weyl-Heisenberg operators, their use as error operators and for the construction of Bell states are introduced in Sec.~\ref{sec:systemstates} together with related state families used in this analysis. The stabilizer theory that FIMAX is built upon is revised in Sec.~\ref{sec:stabilizer}. Finally, Sec.~\ref{sec:FIMAX} contains the precise FIMAX protocol and provides an example for its application. For more information, the reader is referred to Ref.~\cite{popp_novel_2024}. 
\section{Results}
\label{sec:results}
FIMAX was shown to perform entanglement distillation with high efficiency compared to other recurrence-based protocols and can distill strongly mixed states, for which other protocols fail \cite{popp_novel_2024}. In this section, we analyze the performance of the described protocol with focus on the share of distillable states for wider classes of states. 
It is known that any $d^2$-dimensional bipartite state $\rho_{in}$ with fidelity $\mathcal{F}_{in} > \frac{1}{d}$ can be distilled by recurrent two-copy distillation, e.g., by the protocol given in Ref.~\cite{horodecki_reduction_1999}. On the other hand, any entangled state with positive partial transposition (PPT) is bound entangled, i.e., cannot be distilled by any protocol. In this contribution, we thus restrict the analyses to states with negative partial transposition (NPT) and low fidelity. We apply FIMAX to bipartite systems of dimension $d=2$ and $d=3$, and compare the share of FIMAX-distillable states to the shares of other two-copy recurrence-based protocols.\\ 
For this purpose, we sample random states in several state families (cf. Sec.~\ref{sec:systemstates} for details).
The first analyzed state family is the set of pure bipartite states, which we sample uniformly according to the Haar measure, serving a natural measure on the space of pure states \cite{zyczkowski_volume_1998}. 
As a second family, we consider uniformly distributed mixtures of the standard Bell basis states $\lbrace |\Omega_{k,l} \rangle \rbrace$, constructed via the Weyl operators according to \eqref{eq:bellStates} in Sec.~\ref{sec:bellstates}. These Bell-diagonal states (BDS) of the form \eqref{eq:bellDiagonalInputStates} have a special entanglement structure, which is related to the properties of the Weyl operators.
Finally, we analyze uniformly distributed mixtures of different Bell states, called ``generalized Bell states'', which are defined in Sec.~\ref{sec:gBDS}. Similar to the standard Bell basis, these generalized Bell states are also maximally entangled states that form a basis and can be constructed via so-called ``generalized Weyl operators'' as in \eqref{eq:gen_weyl_ops}. However, due to additional phases, the strong algebraic relations between the Bell states are generally lost \cite{popp_special_2024}. \\
To focus on low-fidelity states, we impose one of two restrictions on the set of sampled NPT states. In Sec.~\ref{sec:normalfid}, we require the fidelity $\mathcal{F}_{in} = \langle \Omega_{0,0}| \rho_{in}|\Omega_{0,0}\rangle$ with the specific target state $|\Omega_{0,0}\rangle$ to be smaller than $\frac{1}{d}$. As a stricter condition for low-fidelity, in Sec.~\ref{sec:strictfid}, we require $\langle \Omega_{k,l}| \rho_{in} | \Omega_{k,l} \rangle  \leq \frac{1}{d}$ for all $k,l$. The latter means, that the fidelity of the input state with any of the maximally entangled Bell states \eqref{eq:bellStates} is less than the critical value for certain two-copy distillability. \\
For the families of pure states and standard Bell-diagonal states (BDS), we sample uniformly distributed states and filter this set to $10,000$ NPT low-fidelity states. For the generalized Bell-diagonal states (gBDS), we construct $1,000$ generalized Bell-bases and for each sample $100$ diagonal NPT states with low-fidelity. For these states, the shares of distillable states are then evaluated. Here, we present maximum, minimum, mean and median share of the $1,000$ systems corresponding to the different generalized Bell bases. Note that due to the limited sample size of $100$ states, the sample statistics have limited accuracy, especially for the max and min values. \\   
The share of distillable states of this set is then determined for the following recurrent two-copy protocols: FIMAX \cite{popp_novel_2024}, ADGJ \cite{alber_efficient_2001}, DEJMPS \cite{deutsch_quantum_1996} ($d=2$ only), P1-P2 \cite{miguel-ramiro_efficient_2018} and BBPSSW in its generalization to $d \geq 2$ \cite{bennett_purification_1996, horodecki_reduction_1999}.
\subsection{Distillability of low-fidelity states}
\label{sec:normalfid}
\begin{figure}[H]
    \hspace{-1em}
    \includegraphics[width=0.5\textwidth]{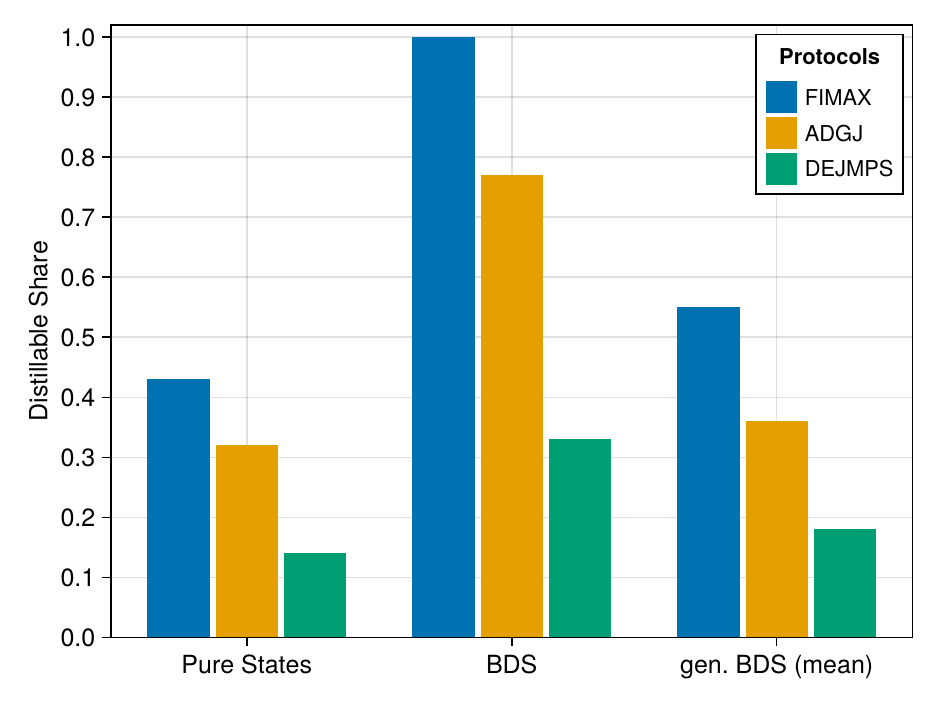}
    \caption{Share of distillable states for $d=2$ and $\mathcal{F}_{in} \leq \frac{1}{2}.$}
    \label{fig:dist2}
\end{figure}

\begin{figure}[H]
    \hspace{-1em}
    \includegraphics[width=0.5\textwidth]{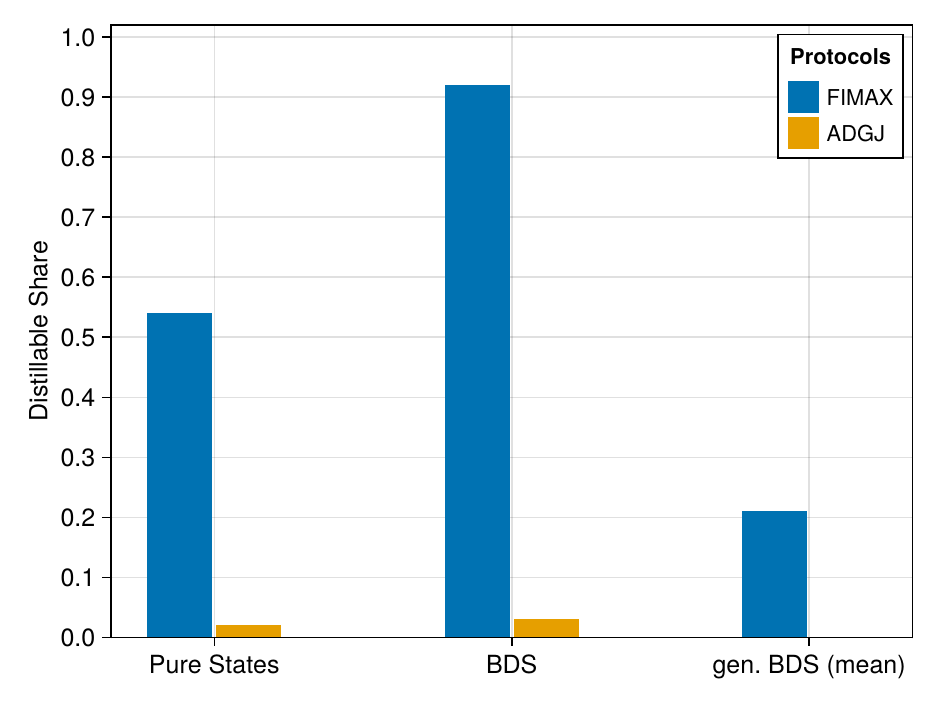}
    \caption{Share of distillable states for $d=3$ and $\mathcal{F}_{in} \leq \frac{1}{3}.$}
    \label{fig:dist3}
\end{figure}

First, consider NPT states for $d=2$ that fulfill $\mathcal{F}_{in} \leq \frac{1}{2}$. The results are summarized in Figure~\ref{fig:dist2}.
For the family of pure bipartite states, FIMAX achieves the highest share of distillable states  with $43 \%$ of analyzed low-fidelity states, followed by ADGJ with $32\%$ and DEJMPS with $14\%$. P1-P2 and BBPSSW cannot distill any such states.
In the set of standard Bell-diagonal states,
FIMAX can distill all states, ADGJ $77 \%$, DEJMPS $33 \%$, P1-P2 $0 \%$ and also BBPSSW $0 \%$.
Considering generalized Bell-diagonal states, the distillable shares are again nonzero only for FIMAX (max: $100 \%$, min: $0\%$, mean: $55\%$, median: $55\%$), ADGJ (max: $85 \%$, min: $0\%$, mean: $36\%$, median: $35\%$ and DEJMPS (max: $43 \%$, min: $0\%$, mean: $18\%$, median: $20\%$). \\
 The equivalent analysis for $d=3$ is shown in Figure \ref{fig:dist3}. FIMAX can distill $54 \%$ of uniformly sampled pure states sufficing the restrictions of NPT and $\mathcal{F}_{in}\leq \frac{1}{3}$. ADGJ achieves a rate of $2 \%$, while P1-P2 and BBPSSW cannot distill any of the sampled states.
Considering the standard Bell-diagonal states, the distillable shares are as follows: FIMAX: $92 \%$,  ADGJ: $3 \%$, P1-P2: $0 \%$, BBPSSW: $0 \%$.
For generalized Bell-diagonal states with $d=3$ and $\mathcal{F}_{in} \leq \frac{1}{3}$, the results for the maximal, minimal, mean and median share among created generalized Bell-diagonal systems are the following: FIMAX: (max: $54 \%$, min: $< 1\%$, mean: $21\%$, median: $21\%$), ADGJ: (max: $3 \%$, min: $0\%$, mean: $< 1\%$, median: $< 1 \%$). The other protocols (P1-P2, BBPSSW) cannot distill any states in all systems.
\subsection{Distillability of strictly low-fidelity states}
\label{sec:strictfid}
\begin{figure}[H]
    \hspace{-0.5em}
    \includegraphics[width=0.5\textwidth]{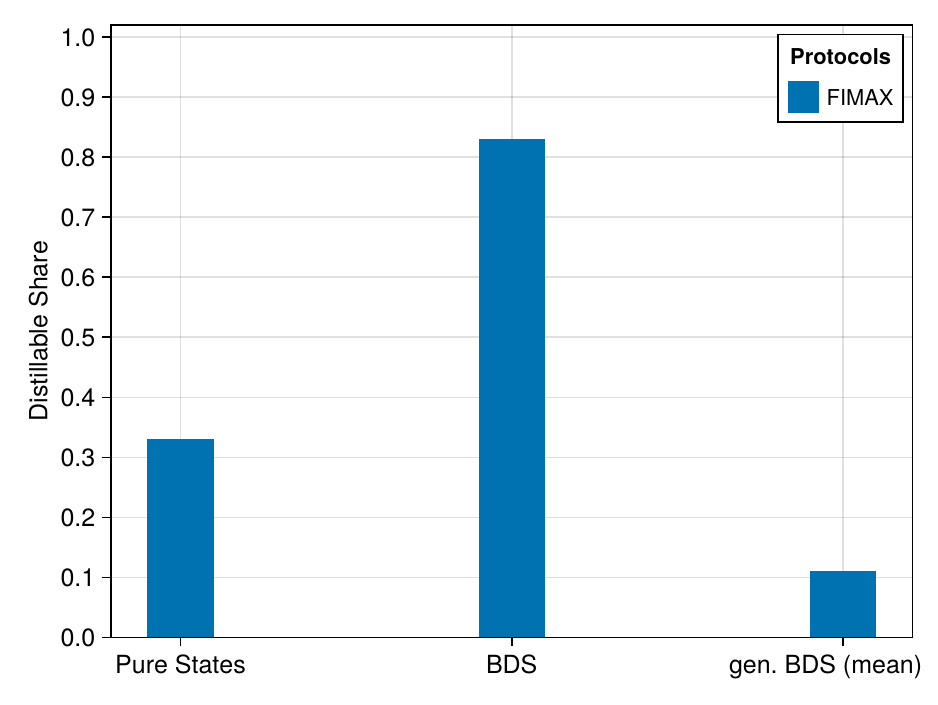}
    \caption{Share of distillable states for $d=3$ and $\langle \Omega_{k,l} | \rho_{in} | \Omega_{k,l} \rangle \leq \frac{1}{3}$ for all $k,l$.}
    \label{fig:dist3s}
\end{figure}

Imposing the stricter fidelity restriction $\langle \Omega_{k,l}| \rho_{in} | \Omega_{k,l} \rangle  \leq \frac{1}{d}$ for all Bell states, none of the protocols can distill any state for $d=2$.
For FIMAX, BBPSSW and P12 this is expected, as any state whose fidelities with any Bell state is smaller than $\frac{1}{2}$ is PPT after being projected to Bell-diagonal form, and therefore undistillable. \\
For $d=3$, the results are shown in Figure \ref{fig:dist3s}. FIMAX is the only protocol to distill a significant share of low-fidelity pure states, with a share of $33 \%$. ADGJ can distill $< 1 \%$, while for  P1-P2 and BBPSSW not a single distillable state is observed.
FIMAX can distill $83 \%$ of low-fidelity standard Bell-diagonal states satisfying the strict fidelity restriction. In contrast to the weaker fidelity restriction, none of the other protocols can distill a significant share of such states, as the share of ADGJ is $< 1\%$ and the shares of other protocols are equal to $0\%$. 
The same holds for generalized Bell-diagonal states, for which FIMAX distills a relevant number of states (max: $33 \%$, min: $1\%$, mean: $11 \%$, median: $9\%$).
\subsection{Additional remarks}
With the given sample sizes, the estimated deviation from the reported numbers for the distillable share is $\leq 1\%$ for equivalent analyses.
In addition, we make some qualitative observations regarding the analysis above. 
First, all states that can be distilled with a different protocol, can also be distilled with FIMAX.
Second, all NPT states that cannot be distilled with FIMAX, result in a PPT output state within the numerical precision after one iteration.
Third, all analyzed states that can be distilled with FIMAX produce an output state that has fidelity $\geq\frac{1}{d}$ within the numerical precision after one iteration. 
Finally, all observed states that are not distillable with FIMAX converge to a Bell-diagonal state that is a fixpoint for recurrent application of the protocol.

\section{Conclusion}
\label{sec:conclusion}
In this work, we conducted a comprehensive comparison of the recently introduced stabilizer-based two-copy bipartite entanglement distillation protocol FIMAX \cite{popp_novel_2024} against other recurrent two-copy protocols, including ADGJ \cite{alber_efficient_2001}, DEJMPS \cite{deutsch_quantum_1996}, P1-P2 \cite{miguel-ramiro_efficient_2018}, and the generalized BBPSSW protocol \cite{horodecki_reduction_1999}. We evaluated their performance across three state families: pure states, standard Bell-diagonal states, and generalized Bell-diagonal states.\\
Our analysis focused on low-fidelity states with negative partial transposition in bipartite systems of dimensions $d = 2$ and $d=3$. We found that FIMAX outperforms all other protocols in terms of the share of distillable states. For states with initial fidelity less than $\frac{1}{d}$, only FIMAX, ADGJ, and DEJMPS could distill entanglement. Notably, for $d=3$, FIMAX's performance surpassed the only other protocol capable of distilling any states by more than an order of magnitude. The substantial share of random pure bipartite states with low fidelity distillable by FIMAX underscores its broad applicability, while its effectiveness with Bell-diagonal states highlights how it leverages the unique algebraic structure of these states. FIMAX's advantage becomes even more pronounced under stricter fidelity constraints. When all elements of the Bell basis have low fidelity with the input state, it stands as the sole considered protocol capable of distilling entanglement, distilling a significant share of states in all analyzed state families. \\
The fact that every state distillable by other protocols was also distillable by FIMAX underlines its exceptional performance in handling low-fidelity states and emphasizes the protocol's robustness.
Importantly, we observed that any state distillable by FIMAX achieves a fidelity greater than $\frac{1}{d}$ after the first iteration, enabling other protocols to further distill these states. Conversely, states undistillable by this protocol become positive under partial transposition (PPT) after the first application, hence becoming inherently undistillable. Leveraging these properties can lead to a resource-efficient application of the protocol, as the distinction of two-copy distillability can principally be made after just one iteration. Our observations provide valuable insights for future research into stabilizer-based approaches for entanglement distillation and the structure of entanglement.\\
In conclusion, combined with its confirmed high efficiency \cite{popp_novel_2024}, our findings demonstrate that FIMAX is very effective for two-copy distillation, particularly in high-noise regimes where other protocols fail. This positions FIMAX as a powerful tool in quantum information processing, potentially accelerating the development of reliable quantum communication in practically relevant, thus noisy, environments.
\section{Methods}
\label{sec:methods}
Finally, we provide details about the analyzed states and systems, the stabilizer-based methods, and the application of the protocol, which we demonstrate by an explicit example.
\subsection{Considered systems and states}
\label{sec:systemstates}
\begin{figure}[H]
    \includegraphics[width=0.35\textwidth]{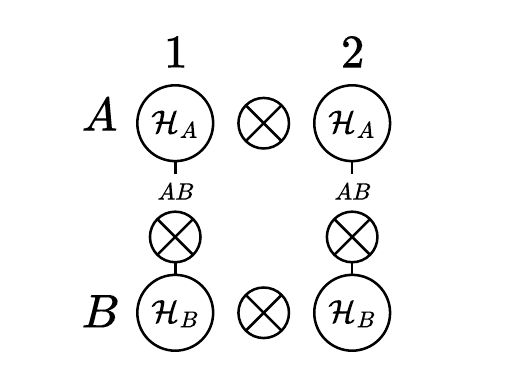}
    \caption{Hilbert space $\hs^{\otimes 2}$ of two-copies of a bipartite system $\hs = \hs_A \otAB \hs_B$.}
    \label{fig:hsTensor}
\end{figure}

Consider the $d^2$-dimensional Hilbert space $\hs \equiv \hs_A \otAB \hs_B \cong \mathbbm{C}^d \otimes \mathbbm{C}^d$ of bipartite quantum states and the corresponding Hilbert space $\hs^{\otimes 2} \equiv \bigotimes_{n=1}^2 \hs \cong \hs_A^{\otimes 2} \otAB \hs_B^{\otimes 2}$ of two-copy bipartite quantum states. 
Figure \ref{fig:hsTensor} visualizes the structure of $\hs^{\otimes 2}$.
In this contribution, we restrict $d$ to be prime. 
$\denop(\hs)$ denotes the set of density operators on $\hs$. 
Further, $\w \equiv \exp(\frac{2 \pi i}{d})$ and $\Zd \equiv \mathbbm{Z}/d\mathbbm{Z}$ denotes the quotient ring of integers with addition and multiplication modulo $d$.
\subsubsection{Weyl errors}
\label{sec:errors}
The \textit{Weyl(-Heisenberg) operators} are a unitary generalization of the two-dimensional Pauli operators to higher dimensions $d$. They are defined as
\begin{flalign}
    \label{eq:weylOperators}    
    W_{k,l} &:= \sum_{j}\w^{j k} ~|j\rangle \langle j+l|,~~j,k,l \in \Zd ,
\end{flalign} 
\noindent and obey the so-called \textit{Weyl-relations}
\begin{flalign}
    \label{eq:weylRelations}
           &W_{k_1,l_1}W_{k_2,l_2} = \w^{l_1 k_2}~W_{k_1+k_2, l_1+l_2},  \\
           &W_{k,l}^\dagger = \w^{k l}~W_{-k, -l} = W_{k,l}^{-1} \notag.
\end{flalign}
The Weyl operators form a so-called ``nice error basis'' \cite{knill_non-binary_1996}, representing both phase and shift errors, corresponding to the indices $k,l$, respectively.
As both copies of the two-copy input state may be affected by errors, we define 
\begin{flalign}
    \label{eq:errorOperators}
    &e := \left[ \begin{pmatrix} \begin{smallmatrix} k_1 \\  l_1 \end{smallmatrix} \end{pmatrix}, \begin{pmatrix} \begin{smallmatrix} k_2 \\ l_2  \end{smallmatrix} \end{pmatrix} \right]  \in \Zd^2 \times \Zd^2, \\
    & W(e) := W_{k_1, l_1} \otimes W_{k_2, l_2} \notag
\end{flalign} 
as a \textit{two-copy error basis}.
The elements $e \in \Zd^2 \times \Zd^2$ are called \textit{error elements}.
Because of the Weyl relations \eqref{eq:weylRelations} and ignoring the for this work irrelevant phases, these two-copy error operators form a group, which naturally extends to the error elements via elementwise addition modulo $d$ by $W(e)W(f) \propto W(e + f)$. This group is denoted by
\begin{flalign}
    \label{eq:errorGroups}
    \E_2 &:= \left\{  W(e) ~|~ e \in \Zd^2 \times \Zd^2 \right\}.
\end{flalign}
\subsubsection{Bell states}
\label{sec:bellstates}
Besides their role as error operators, the Weyl operators can also be used to define a basis of the state space for pure bipartite qudits, whose elements are maximally entangled and called \textit{Bell states}. They are defined as:
\begin{flalign}
    \label{eq:bellStates}
        |\Omega_{0,0}\rangle &:= \frac{1}{\sqrt{d}}\sum_{i} | i \rangle \otAB |i \rangle,~~i \in \Zd, \notag \\
        |\Okl\rangle &:= (W_{k,l} \otAB \mathbbm{1}_d)~|\Omega_{0,0}\rangle, ~~ k,l \in \Zd.        
\end{flalign}
The \textit{two-copy Bell basis} is then defined by the two-copy states
\begin{flalign}
    \label{eq:two-copy_bellbasis}
    | \Omega(e) \rangle &:= |\Omega_{k_1,l_1}\rangle \otAB |\Omega_{k_2,l_2} \rangle, \notag\\
    &\equiv (W(e) \otAB \mathbbm{1}_{d^2})~|\Omega_{0,0}\rangle^{\otimes 2},
\end{flalign}
where we use the notation introduced in \eqref{eq:errorOperators}.
In this basis, any bipartite two-copy state $\rho^{\otimes 2}$ can be then written as:
\begin{flalign}
    \label{eq:inputStates}
    &\rho^{\otimes 2} = \sum_{e,f} \rho(e, f) ~|\Omega(e)\rangle \langle \Omega(f)|, 
\end{flalign}
where $e,f \in \Zd^2 \times \Zd^2$ and we use the notation $\rho(\cdot,\cdot)$ for the elements of the density matrix representing the state in the two-copy Bell basis.
A state $\rho \in \denop(\hs)$ is called Bell-diagonal if it is of the form
\begin{flalign}
    \label{eq:bellDiagonalInputStates}
    \rho = \sum_{k,l} p_{k,l}~|\Omega_{k,l}\rangle\langle \Omega_{k,l}|,
\end{flalign}
where $p_{k,l} \geq 0$ and $\sum_{k,l} p_{k,l} = 1$. In this case, it can be interpreted as a mixture of Bell states \eqref{eq:bellStates}, allowing to generate those states in experiment \cite{hiesmayr_complementarity_2013,}. Due to its algebraic and geometric properties implied by the Weyl-relations and its special entanglement structure, this family of states is also known as ``Magic Simplex'' \cite{baumgartner_special_2007, popp_almost_2022, popp_comparing_2023}. \\
The two-copy state $\rho^{\otimes 2} \in \denop(\hs^{\otimes 2})$ of a Bell-diagonal state takes the following diagonal form:
\begin{flalign}
    \label{eq:twoCopyBellDiagonal}
    \rho^{\otimes 2} &= \sum_{e} \rho(e, e) ~|\Omega(e)\rangle \langle \Omega(e)|.
\end{flalign}
Any state $\rho \in \denop(\hs)$ can be projected to Bell-diagonal form by application of the ``Weyl twirl channel'' \cite{popp_special_2024}.
\begin{flalign}
    \label{eq:weyltwirl}
    \rho \mapsto \frac{1}{d^2}\sum_{i,j} W_{i,j}\otAB W_{-i,j} ~\rho~ (W_{i,j}\otAB W_{-i,j})^\dagger,
\end{flalign}
With $i,j \in \Zd$. Crucially for entanglement distillation, this channel can be realized by LOCC operations.
\subsubsection{Generalized Bell states}
\label{sec:gBDS}
The family of generalized Bell basis states and corresponding ``generalized Bell-diagonal states (gBDS)'' is defined as follows (cf. \cite{popp_special_2024}). Let $(\alpha_{s,t})_{s,t \in \mathbb{Z}_d}$ with $|\alpha_{s,t}|=1~ \forall s,t \in \mathbb{Z}_d$ be a matrix of complex phases. For each $\alpha$, we can define an orthonormal basis of maximally entangled states and corresponding diagonal states:
\begin{flalign}
    \label{eq:gen_weyl_ops}
    &V^\alpha_{k,l} := \sum_j w^{j k} \alpha_{j+l,l}~|j\rangle\langle j+l|, \\
    &|\Phi^\alpha_{k,l} \rangle := (V^\alpha_{k,l} \otimes \mathbbm{1}_d)~|\Omega_{0,0} \rangle, \notag \\
    &\rho = \sum_{k,l} p_{k,l}~|\Phi^\alpha_{k,l} \rangle \langle \Phi^\alpha_{k,l}|, \notag
\end{flalign}
where $j,k,l \in \Zd$, $p_{k,l}\geq 0$ and $\sum_{k,l} p_{k,l}= 1$. The main difference to the standard Bell states is that the generalized Weyl operators $V^\alpha_{k,l}$ do not obey the Weyl relations \eqref{eq:weylRelations}, resulting in a loss of algebraic structure and a change in the entanglement structure of diagonal states.

\subsection{Stabilizer Theory}
\label{sec:stabilizer}
In the following, we shortly introduce the relevant stabilizer-related objects, required to apply the FIMAX distillation protocol. For detailed information about stabilizer theory and its application to entanglement distillation, see, e.g., Refs. ~\cite{popp_novel_2024, matsumoto_conversion_2003, knill_non-binary_1996, rains_nonbinary_1999}. \\
Given the error group of Weyl error operators $\E_2$ as in \eqref{eq:errorGroups}, a \textit{stabilizer} $S$ is an abelian subgroup of $\E_2$. Ref.~\cite{popp_novel_2024} shows that for prime dimensions, any stabilizer relevant for entanglement distillation is generated by one error operator $W(g)$ with a corresponding error element $g \in \Zd^2 \times \Zd^2$. Such stabilizers contain exactly $d$ elements and are of the form $S = \lbrace \mathbbm{1}, W(g), W(2g), \dots ,W((d-1)g) \rbrace$. \\
A stabilizer $S$ and the corresponding generating element $g$ allow partitioning the set of two-copy error elements in two ways that are relevant for the FIMAX protocol.
First, it decomposes into \textit{cosets} of the stabilizer $S$, which are defined as the sets 
\begin{flalign}
    \label{eq:coset}
    &C(e) := \lbrace e, e+g, e+2g, \dots, e + (d-1)g \rbrace,
\end{flalign}
where $e \in \Zd^2 \times \Zd^2$.
Second, a stabilizer implies a decomposition of error elements by their \textit{symplectic product} with the generator 
$g = \left[ \begin{pmatrix} \begin{smallmatrix} a_1 \\  b_1 \end{smallmatrix} \end{pmatrix}, \begin{pmatrix} \begin{smallmatrix} a_2 \\ b_2  \end{smallmatrix} \end{pmatrix} \right]  $. For an error element 
$e = \left[ \begin{pmatrix} \begin{smallmatrix} k_1 \\  l_1 \end{smallmatrix} \end{pmatrix}, \begin{pmatrix} \begin{smallmatrix} k_2 \\ l_2  \end{smallmatrix} \end{pmatrix} \right]  $, it is defined as $\langle g,e \rangle := \sum_{n=1}^2 b_n k_n - a_n l_n$. We have a decomposition of error elements into $d$ subsets of the form
\begin{flalign}
    \label{eq:sErrors}
    \varepsilon(s) :=  \lbrace e \in \Zd^2 \times \Zd^2 ~|~ \langle g, e \rangle = s \rbrace, ~s \in \Zd.
\end{flalign}
Note that both decompositions depend on the generator $g$ and that if $e \in \varepsilon(s)$, then all error elements of the error coset $C(e)$ are in $\varepsilon(s)$, as well. \\
The eigenspaces of the generator $W(g)$ decompose the Hilbert space on which the error operators act in so-called \textit{codespaces} $\mQ$. Using the two-copy Bell basis to represent the input states \eqref{eq:two-copy_bellbasis}, it is clear that $W(e)$ acts on the two parts of the $A$-side of the bipartite system. Denoting the eigenspaces of $W(g)$ with eigenvalue corresponding to $x$ by $\mQ(x)$, we have $\hs_A^{\otimes 2} \cong \bigoplus_x \mQ(x)$. 
For nontrivial generators, the codespaces are of dimension $d$ and can be spanned by an orthonormal basis of eigenvectors of $W(g)$, the  so-called \textit{codewords} $\lbrace |u_{x,k} \rangle \rbrace_{x,k \in \Zd}$.
A unitary map $U: |x\rangle \otimes |k\rangle \mapsto |u_{x,k}\rangle \in \mQ(x) \subset \hs_A^{\otimes 2}$ is called \textit{encoding} for $S$.
Ref.~\cite{popp_novel_2024} proposes a \textit{canonic encoding} $U_c$, based on the product eigenvectors of the given generator $W(g) = W_{a_1,b_1} \otimes W_{a_2, b_2}$. $U_c$ has special properties that are leveraged in FIMAX. 
A \textit{stabilizer measurement} performed on $\hs_A^{\otimes 2}$ corresponds to a projection on a certain codespace $\mQ(x)$ and can be expressed as the projection-valued measure $\lbrace \mP(x) \rbrace_{x\in \Zd}$, $\mP(x) = \sum_k | u_{x,k} \rangle \langle u_{x,k}|$. Given $\rho$, the unnormalized post-measurement state for measurement outcome $x$ is $\mP(x)~\rho~\mP(x) \in \mQ(x)$.  \\
Note that the objects defined above relate to the $A$ side of the bipartition, as the elements of $S$ are operators acting on $\hs_A^{\otimes 2}$. To use stabilizer codes for distillation, we also need the corresponding objects for the other side $B$. Denote by $S^\star$ the stabilizer with complex conjugated elements of $S$, generated by $W(g)^\star$ and implying a decomposition $\hs_B^{\otimes 2} \cong \bigoplus_y \mQ^\star(y)$. Then, $U^\star: |y\rangle \otimes |l\rangle \mapsto |u_{y,l}^\star \rangle \in \mQ^\star(y)$ defines an encoding for $S^\star$ and $\mP^\star(y) =  \sum_l | u_{y,l}^\star \rangle \langle u_{y,l}^\star|$ defines the projections for the stabilizer measurements corresponding to $S^\star$.

\subsection{FIMAX}
Having introduced the relevant objects in the previous section, we now specify the two-copy recurrence protocol FIMAX and provide an example for its application.
\label{sec:FIMAX}
\subsubsection{Protocol}
\label{sec:protocol}
Let $\rho_{in} \in \denop(\hs)$ be a bipartite state in prime dimension $d$ and $\rho_{in}^{\otimes 2} \in \denop(\hs^{\otimes 2})$ be the two-copy state as in \eqref{eq:inputStates}. Define the probability distribution $\prob(e) := \rho_{in}^{\otimes 2}(e,e)$ and denote corresponding probabilities by $\Prob$. 
\begin{enumerate}[left=0pt]
    \item If $\rho_{in}$ is not of Bell-diagonal form \eqref{eq:bellDiagonalInputStates}, apply the Weyl twirl channel \eqref{eq:weyltwirl} to both copies.
    \item For each stabilizer $S$ with generating element $g \in \Zd^2\times \Zd^2$:
    \begin{enumerate}[label=(\roman*),left=-15pt]
        \item Partition the error elements $e$ according to their symplectic product \eqref{eq:sErrors} and calculate $\Prob(\varepsilon(s))$. 
        \item For each $s$ with $\Prob(\varepsilon(s)) > 0$ and for each error coset $C = C(e)$, $e \in \varepsilon(s)$ \eqref{eq:coset}, determine $(C_{max},s_{max}) :=\arg\max \frac{\Prob(C)}{\Prob(\varepsilon(s))}$.
    \end{enumerate}
    \item Alice chooses the stabilizer $S_{max}$, maximizing $\frac{\Prob(C_{max})}{\Prob(\varepsilon(s_{max}))}$ among stabilizers, and communicates her choice classically to Bob.
    \item Alice and Bob perform stabilizer measurements for $S_{max}$ and $S_{max}^\star$ with measurement outcomes $a, b$, respectively.
    \item Bob sends $b$ to Alice. Alice declares failure of the protocol if $s_{max} \neq a-b $.
    \item Alice and Bob apply the inverse of the canonic encoding $U_c$ and $U_c^\star$ for $S_{max}$ and $S^\star_{max}$ to their qudits, respectively.
    \item Alice and Bob discard their first qudit containing the measurement outcome.
    \item Alice applies $W^\dagger_{k_{max},l_{max}}$ to the remaining qudit, where $(k_{max},l_{max})$ is determined from $(C_{max},s_{max})$ as described in Ref.\cite{popp_novel_2024}. 
\end{enumerate}
Note that FIMAX maps Bell-diagonal states to Bell-diagonal states, so the first step of the protocol is only required once for recurrent application of the protocol.\\
\subsubsection{Example}
\label{sec:example}
As an explicit example to better understand the relevant stabilizer-related objects leveraged by FIMAX, consider the case $d=3$ and the Bell-diagonal input state, taking the form in the basis \eqref{eq:bellStates}, ordered  first by shift then by phase index as $|\Omega_{0,0} \rangle, |\Omega_{1,0} \rangle, \cdots, |\Omega_{1,2} \rangle,|\Omega_{2,2} \rangle $:
\begin{flalign}
    \rho_{in} = \begin{pmatrix} \begin{smallmatrix} 
      0.06 &   &   &   &   &   &   &   &  \\ 
      & 0.06 &   &   &   &   &   &   &  \\
      &   & 0.06 &   &   &   &   &   &  \\
      &   &   & 0.06 &   &   &   &   &  \\
      &   &   &   & 0.06 &   &   &   &  \\
      &   &   &   &   & \textbf{0.56} &   &   &  \\
      &   &   &   &   &   & 0.06 &   &  \\
      &   &   &   &   &   &   & 0.06 &  \\
      &   &   &   &   &   &   &   & 0.06 \\
     \end{smallmatrix} \end{pmatrix} \notag
\end{flalign}
This state has small fidelity $\mathcal{F}=0.06$, but a large overlap of $0.56$ with the Bell state $|\Omega_{2,1}\rangle$.
Comparing the coset probabilities for all stabilizers according to step 2 of FIMAX, the stabilizer with the generating element 
$g = \left[ \begin{pmatrix} \begin{smallmatrix} 1 \\  0 \end{smallmatrix} \end{pmatrix}, \begin{pmatrix} \begin{smallmatrix} 1 \\ 0  \end{smallmatrix} \end{pmatrix} \right]  \in \mathcal{Z}^2_3 \times \mathcal{Z}^2_3$ implies the best partition of error elements. More precisely,  $g$ corresponds to the operator $W(g) = W_{1,0}\otimes W_{1,0}$ which generates the stabilizer 
\begin{flalign}
    S_{max} &= \lbrace W(0), W(g), W(2g) \rbrace \\
    &= \lbrace \mathbbm{1}_3 \otimes \mathbbm{1}_3, W_{1,0} \otimes W_{1,0}, W_{2,0}\otimes W_{2,0} \rbrace. \notag
\end{flalign} 
The coset 
\begin{flalign}
    &~~~~C_{max} = \left\{
    \left[ \begin{pmatrix} \begin{smallmatrix} 0 \\  1 \end{smallmatrix} \end{pmatrix}, \begin{pmatrix} \begin{smallmatrix} 0 \\ 1  \end{smallmatrix} \end{pmatrix} \right],  
    \left[ \begin{pmatrix} \begin{smallmatrix} 1 \\  1 \end{smallmatrix} \end{pmatrix}, \begin{pmatrix} \begin{smallmatrix} 1 \\ 1  \end{smallmatrix} \end{pmatrix} \right] , 
    \left[ \begin{pmatrix} \begin{smallmatrix} 2 \\  1 \end{smallmatrix} \end{pmatrix}, \begin{pmatrix} \begin{smallmatrix} 2 \\ 1  \end{smallmatrix} \end{pmatrix} \right]
    \right\}&&  \notag
\end{flalign}
is a subset of $\varepsilon(1)$, i.e., all error elements having symplectic product $\langle g, e \rangle = 1 = s_{max}$. 
Based on the probabilities $\Prob$ induced by $\rho_{in}^{\otimes 2}$, $\Prob(\varepsilon(1)) = 0.5$ and $\Prob(C) = 0.315$, which correspond to the larges value of $\frac{\Prob(C)}{\Prob(\varepsilon(s))}$ for all stabilizers $S$, cosets $C$ and values of $s$.\\
The eigenvalues of the generator $W(g)$ are $\omega^x,~x\in \mathcal{Z}_3$, each threefold degenerate. It can be checked that the unitary
\begin{flalign}
    U_c := | x \rangle \otimes | k \rangle \mapsto |u_{x,k} \rangle := |k\rangle \otimes |x-k\rangle
\end{flalign}
maps any computational basis state $| x \rangle \otimes | k \rangle$ to a codeword $|u_{x,k} \rangle$ of the codespace $\mQ(x)$ corresponding to the eigenvalue $\omega^x$ and is therefore an encoding for $S$. In fact, this encoding is the canonic encoding $U_c$ for $S$. Note that in this example, we have $S = S^\star$ and therefore $U_c = U_c^\star$.\\
Performing the stabilizer measurements having outcomes $a, b$ with $a-b=s_{max} = 1$,  according to steps 4 and 5, followed by the application of $U_c^\dagger \otimes U_c^\dagger$ in step 6 yields the following state of the remaining qudits:
\begin{flalign}
    \rho_{out}^\prime = \begin{pmatrix} \begin{smallmatrix} 
      0.02 &   &   &   &   &   &   &   &  \\ 
      & 0.02 &   &   &   &   &   &   &  \\
      &   & 0.02 &   &   &   &   &   &  \\
      &   &   & \textbf{0.63} &   &   &   &   &  \\
      &   &   &   & \textbf{0.13} &   &   &   &  \\
      &   &   &   &   & \textbf{0.13} &   &   &  \\
      &   &   &   &   &   & 0.02 &   &  \\
      &   &   &   &   &   &   & 0.02 &  \\
      &   &   &   &   &   &   &   & 0.02 \\
     \end{smallmatrix} \end{pmatrix} \notag
\end{flalign}
Applying the correction operation $W_{0,2} \otAB \mathbbm{1}_3$ results in the final output state
\begin{flalign}
    \rho_{out} = \begin{pmatrix} \begin{smallmatrix} 
      \textbf{0.63} &   &   &   &   &   &   &   &  \\ 
      & \textbf{0.13} &   &   &   &   &   &   &  \\
      &   & \textbf{0.13} &   &   &   &   &   &  \\
      &   &   & 0.02 &   &   &   &   &  \\
      &   &   &   & 0.02 &   &   &   &  \\
      &   &   &   &   & 0.02 &   &   &  \\
      &   &   &   &   &   & 0.02 &   &  \\
      &   &   &   &   &   &   & 0.02 &  \\
      &   &   &   &   &   &   &   & 0.02 \\
     \end{smallmatrix} \end{pmatrix}. \notag
\end{flalign}
The state $\rho_{out}$ has a higher fidelity and maximal overlap with any Bell state ($\mathcal{F}=0.63$) compared to the input state $\rho_{in}$.
\printbibliography    
\section*{Acknowledgments}
C.P. and B.C.H. acknowledge gratefully that this research was funded in whole, or in part, by the Austrian Science Fund (FWF) project P36102-N (Grant DOI:  10.55776/P36102). For the purpose of open access, the author has applied a CC BY public copyright license to any Author Accepted Manuscript version arising from this submission. The funder played no role in study design, data collection, analysis, and interpretation of data, or the writing of this manuscript.

\section*{Author Contributions Statement}
C.P. developed the methods, carried out the analyses, implemented the software and edited the manuscript. B.C.H. and T.C.S. revised the analyses and proposed improvements. 

\section*{Data availability statement}
All datasets generated during the study are available from the corresponding author on reasonable request.

\section*{Code availability statement}
The software used to generate the reported results is published as a repository and registered open source package ``BellDiagonalQudits.jl'' \cite{popp_belldiagonalqudits_2023}.



\end{multicols}

\end{document}